\documentclass[pre,twocolumn,aps,floatfix]{revtex4}

\usepackage{graphicx}

\begin{document}

\title{
Diffusion on a solid surface: Anomalous is normal}

\author{
J.M. Sancho$^{1}$, A.M. Lacasta$^{2}$,
K. Lindenberg$^{3}$, I. M.  Sokolov$^{4}$ and A.H. Romero$^{5}$ 
}
\affiliation{
$^{(1)}$
Departament d'Estructura i Constituents de la Mat\`eria,
Facultat de F\'{\i}sica, Universitat de Barcelona,
Diagonal 647, E-08028 Barcelona, Spain\\
$^{(2)}$
Departament de F\'{\i}sica Aplicada,
Universitat Polit\`{e}cnica de Catalunya
Avinguda Dr. Mara\~{n}on 44, E-08028 Barcelona, Spain.\\
$^{(3)}$
Department of Chemistry and Biochemistry 0340, and Institute for Nonlinear 
Science,
University of California, San Diego, La Jolla, California 92093-0340.USA.\\
$^{(4)}$
Institut f\"{u}r Physik, Humboldt Universit{\ae}t zu Berlin, Newtonstr. 15, 
12489 Berlin, Germany.\\
$^{(5)}$
Advanced Materials Department, IPICyT, Camino a la presa San
Jos\'e 2055,CP 78216, San Luis Potos\'\i, SLP, M\'exico.
}

\date{\today}

\begin{abstract}

We present a numerical study of classical particles 
diffusing on a solid surface. The particles' motion is modeled by an  underdamped 
Langevin equation with ordinary thermal noise. The particle-surface
interaction is described by a periodic or a random two
dimensional potential.  The model leads to
a rich variety of different transport regimes, some of which 
correspond to anomalous diffusion such as has recently been observed
in experiments and Monte Carlo simulations.
We show that this anomalous behavior is controlled by the friction 
coefficient, and stress that it emerges  naturally in a system
described by ordinary canonical Maxwell-Boltzmann statistics.

\end{abstract}

\maketitle

PACS: 05.40-a, 68.43.Jk, 68.35.Fx

%\maketitle

%\twocolumn
%\narrowtext

Diffusion of atoms, molecules, and clusters on solid surfaces
occurs in a number of modern technologies involving
self-assembled molecular film growth,
catalysis, and surface-bound nanostructures~\cite{nanoala}.
The study of the motion of small and large
organic molecules~\cite{phd,prl1}, and of adsorbed metal clusters composed
of tens and even hundreds of atoms~\cite{phystoday,prl2}, has led to the
unexpected observation that, as with single atoms~\cite{oneatom},
long jumps may play a dominant role in these motions.

Theoretical, numerical, and phenomenological discussions of surface diffusion
have led to the clear understanding that jumps beyond nearest neighbors are
ubiquitous in some parameter regimes~\cite{theory}. However, these studies
focus on the fact that the motion is necessarily diffusive (which is the case at
very long time scales).  The possibility that jumps can be so
long as to lead to superdiffusive motion over appreciable intermediate
time scales is recognized as an interesting problem, but one in which
L\'{e}vy  walks or flights~\cite{mike} are invoked as a model input.
AlthoughL\'{e}vy-walk-like behavior is clearly observed in Hamiltonian
systems~\cite{Klafterb,vega} and in microcanonical
simulations~\cite{vega,Guantes},
it is generally believed that such a fine signature
of chaos is fully smeared away by thermal fluctuations.
In the present work we show that L\'{e}vy-like statistics appear quite
naturally over long time scales
within the usual Langevin framework for underdamped
motion in a periodic or a random potential.

While a detailed analysis of surface diffusion requires
extensive calculations (e.g., {\em ab initio}, or molecular dynamics),
even the most powerful currently available computers
can not carry such calculations to anywhere near experimentally relevant
time scales~\cite{gross}.  Moreover, current experimental
probes of the topography of surfaces, scanning tunneling
microscopy and atomic force microscopy,
are usually carried out at relatively high temperatures, which leads to
additional difficulties for first-principles calculations.
Therefore, simpler approaches are essential and valuable~\cite{theory}.

We consider a generic model of classical particles
moving in a two-dimensional potential, under the action of thermal
fluctuations and dissipation, the important control parameter 
being the friction coefficient.  In spite of the simplicity of the
model, we find that it is able to reproduce the entire range of
experimentally and computationally observed phenomena, 
from superdiffusion all the way to subdiffusion.

The equation of motion of a particle of mass $m$ on the surface is
\begin{equation}
m{\ddot {\bf x}} = - \nabla V({\bf x}/\lambda) - \mu {\dot {\bf x}} + {\bf \xi}(t)
\label{eq:1}
\end{equation}
where $\lambda$ is the characteristic length scale of the potential.
The parameter $\mu$ is the coefficient of friction, and the
$\xi_i(t)$ are mutually uncorrelated white noises that obey the
fluctuation-dissipation relation
$\langle \xi_i(t)\xi_j(t')\rangle =2\mu k_BT\delta_{ij}\delta(t-t').
$
We first consider the nonseparable periodic potential
\begin{equation}
V(x,y)=V_0 \cos\left(\frac{\pi x}{\lambda} + \frac{\pi y}{\lambda}\right)
\;\cos\left(\frac{\pi x}{\lambda} - \frac{\pi y}{\lambda}\right),
\end{equation}
where $V_0$ is the barrier height at the saddle points. 

Equation~(\ref{eq:1}) can be rewritten in scaled dimensionless
variables, $r_x=x/\lambda$, $r_y=y/\lambda$, and
$s=\sqrt{V_0/m\lambda} t$, leaving only two independent
parameters, the scaled temperature ${\mathcal T}$ and
the scaled dissipation $\gamma$,
\begin{equation}
{\mathcal T}= k_BT/V_0, \qquad
\gamma=\mu\lambda/\sqrt{mV_0}.
\end{equation}
We study 
four properties of the motion of the particle: the mean square displacement,
the dependence of the diffusion coefficient on friction,
the probability density function of displacements, and the velocity
power spectrum.

\begin{figure}
\begin{center}
\includegraphics[width = 9cm]{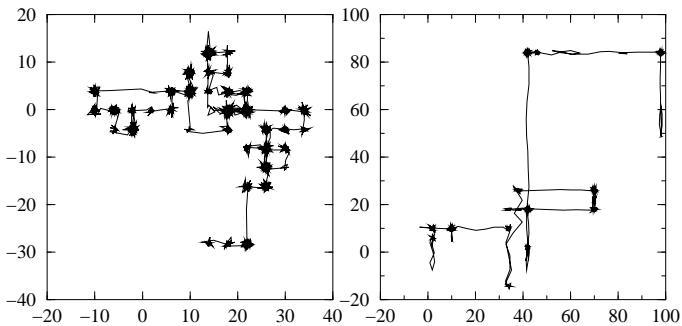}
\end{center}
\caption{
Left: A trajectory for $\gamma=1$ over $t=20,000$ time units.
Right: A trajectory for $\gamma=0.04$ over $t=15,000$ time
units. The period of the potential is $\lambda=4$. Note the different 
scales in the two panels. 
}
\label{fig2}
\end{figure}

Normal diffusive behavior is characterized by a linear time
dependence of the mean square displacement, $\langle r^2(s)\rangle \sim s$. 
Non-diffusive behavior shows a different time
dependence, $\langle r^2(s)\rangle \sim s^\alpha$,
with $\alpha>1$ ($<1$) for superdiffusive (subdiffusive) motion.
In Fig.~\ref{fig2} we show typical trajectories obtained for
two friction coefficients upon numerical simulation of the 
equations of motion with ${\cal T} =0.2$ (we use this value throughout).
One (left panel) is for a large friction coefficient,
and the particle follows typical diffusive
motion characterized by short steps of length $\sim \lambda$ and frequent changes
in direction.  The other 
(right panel) corresponds to a small friction coefficient and clearly shows the
preponderance of long ($\gg \lambda$) tracks along cartesian
coordinates.

\begin{figure}
\begin{center}
\includegraphics[width = 7.0cm]{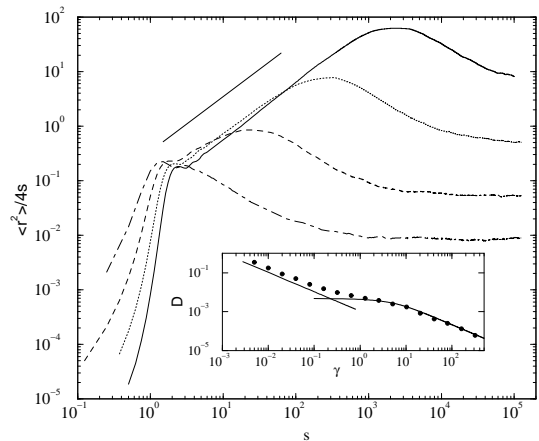}
\end{center}
\caption{
$\langle r^2\rangle/4s$ for a particle in the periodic potential,
for $\gamma=0.0004$ (solid), $0.004$ (dotted),
$0.04$ (dashed) and $0.4$ (dot-dashed).  
The straight-line segment has unit slope as a guide to the eye.
Inset: 
Diffusion coefficient as a function of $\gamma$.  
The solid lines correspond to the theoretical calculations 
(Eqs.\ (\ref{cal1})-(\ref{cal2})).}
\label{fig3}
\end{figure}

The evolution of
$\langle r^2 \rangle$ averaged over $1000$ particles
is shown in Fig.~\ref{fig3} for several friction coefficients.
For very long times the motion is diffusive, 
as expected, but for small $\gamma$ and
at intermediate times there is clear \emph{superdiffusive}
ballistic ($\alpha=2$) behavior
over several decades in time.
This is reflective of the long straight
stretches seen in the low-$\gamma$ trajectory in
Fig.~\ref{fig2}. We stress that this behavior has emerged
naturally and has not required explicit insertion of any but 
ordinary thermal fluctuations in the model.

Eventhough the motion of the particle may include long
superdiffusive stretches, at long times the motion is
necessarily diffusive.  
The dependence of the diffusion coefficient on the friction
for small and for large $\gamma$ 
can be obtained analytically using the approximate relation
$D\approx \langle l^2 \rangle /2\tau$, where $\langle l^2\rangle$ is the
mean square size of a jump out of one well and into
another, and $\tau^{-1}$ is the 
mean jump rate (related to the familiar ``mean escape rate''). 
In the overdamped regime, 
jumps typically occur from one well to a neighboring well, so
$\langle l^2\rangle \approx 1$. Familiar Kramers formulas
can be used to obtain the mean escape rate~\cite{Hanggi}, with the result
\begin{equation}
D \sim \frac{1}{2\pi}
\left( \sqrt{\frac{\gamma^2}{4}+2\pi^2}-\frac{\gamma}{2}
\right ) e^{-\frac{1}{{\cal T}}}
\sim \frac{\pi}{\gamma}  e^{-\frac{1}{{\cal T}}}.
\label{cal1}
\end{equation}
The $\gamma^{-1}$ dependence of $D$ arises because $\langle
l^2\rangle$ is independent of $\gamma$ while $\tau\sim \gamma$.
In the underdamped limit, standard results~\cite{Hanggi} lead to 
$\tau \sim 1/\gamma$~\cite{sancho}, but now
$\langle l^2\rangle \sim \gamma^{-2}$~\cite{we}.
One obtains 
\begin{equation}
D\sim \frac{\pi{\cal T}}{4\gamma} 
e^{-\frac{1}{{\cal T}}},
\label{cal2}
\end{equation}
that is, again an inverse dependence on friction. 
The theoretical diffusion coefficient as a function of the friction
parameter is shown as the solid curves in the inset of 
Fig.~\ref{fig3}.  The symbols are the simulation results.
The $\gamma^{-1}$ dependences have been noted in the
literature~\cite{theory}, but we have provided explicit forms with
\emph{no adjustable parameters}.

\begin{figure}
\begin{center}
\includegraphics[width = 8cm]{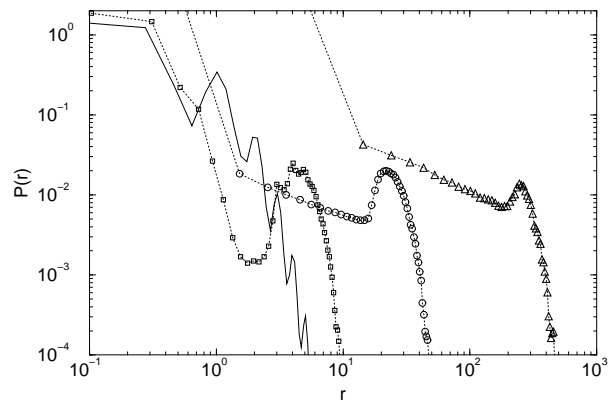}
\end{center}
\caption{
Log-log plot of probability distribution functions for the particle
displacement for $\gamma=0.0004$ and three different values of 
time intervals : $\tau_s= 20$ (squares), 
$100$ (circles) and $1000$ (triangles), and for
$\gamma=1.0$ at the time interval  $\tau_s= 100$ (solid curve).  
}
\label{pdf}
\end{figure}

The diffusion coefficient characterizes only the very long
time asymptotic dynamics.
The interesting intermediate dynamics in the low
friction regime that gives rise to long stretches of ballistic motion 
is reflected in the probability distribution function (pdf) of
particle displacements $r$.
This pdf is shown in Fig.~\ref{pdf} for $\gamma=0.0004$ and
tree different time intervals $\tau_s$.  For comparison, we
also show a typical pdf for high damping ($\gamma=1$) at the
intermediate time interval.
In the high-$\gamma$ curve the highest maximum
corresponds to no jumps (by far the most likely event at short times).  
The next is associated with jumps to a
nearest neighbor well, and so on.  In contrast,
the low-$\gamma$ curves show a 
very different behavior, with features
strongly resembling those of a L\'evy-walk model~\cite{Klafterb,Klaftera}:
a peak at small displacements, a power-law intermediate regime,
and a side hump at high displacements.  Each of these is a
distinct signature of L\'evy-walk-like dynamics, but one must be cautious
in the detailed interpretation of these components.  The persistent small
displacement peak is associated with long trapping periods during which
a particle does not move at all because its energy is not sufficient
to overcome the barrier.  The high displacement peak, which
moves outward with velocity of order unity, is associated
with ballistic motion of those particles that
acquire enough energy to move (and lose it very slowly). 
Genuine L\'evy-walk dynamics also exhibit a low displacement peak
and a superdiffusive peak separated by a power law behavior, 
but there are some important differences.
First, our distribution reflects ballistic transport in the intermediate
regime (in the language of Ref.~\cite{Klaftera}, ballistic
transport occurs when $0<\alpha<1$ in the L\'evy model), whereas 
the regime where the L\'evy model shows the features we have described
is associated with sub-ballistic (but still superdiffusive) behavior
(again, in the language of Ref.~\cite{Klaftera}, the behavior when
$1<\alpha<2$).  Second, the slope in our power law regime 
(approximately $0.7$) is not related to the exponent $\alpha$ 
in the mean square displacement as it is for the L\'evy walk (where the
slope is $4-\alpha$).
Third, our side hump is strongly broadened whereas the
side hump in the L\'evy-walk model is associated with motion at a
single constant velocity.  In our case the 
velocity varies according to the equilibrium 
Maxwell-Boltzmann distribution. Nevertheless, the qualitative features of
our distribution track those of the L\'evy walk.  Note that
the existence of the pronounced side 
hump reflects the fact that the particles performing long steps 
(``flights'') are those with a velocity in the tail of the
Maxwellian distribution.
\begin{figure}
\begin{center}
\includegraphics[width = 8cm]{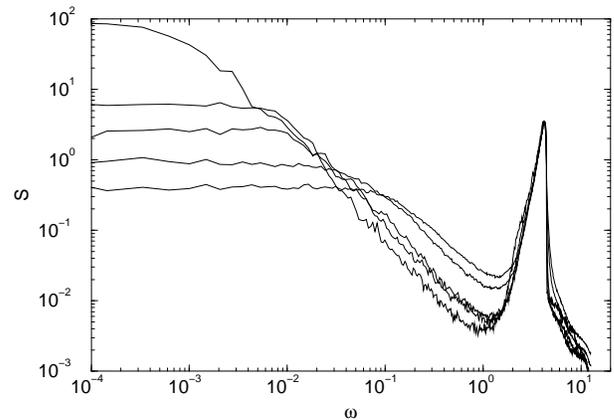}
\end{center}
\caption{$S(\omega)$ for $\gamma=0.0004, 0.004$ and $0.04$ in decreasing
order on the ordinate.  
}
\label{Somega}
\end{figure}

Long ballistic excursions imply velocity correlations
over considerable time intervals.  The velocity power spectrum
$S(\omega )=\left\langle \mathbf{v}(\omega )\mathbf{v}(-\omega )\right\rangle$
for different
values of $\gamma$ is shown in Fig.~\ref{Somega}.  The pronounced
peak at $\omega_0 = \pi \sqrt{2}$ is associated with small
oscillations in one well.  At
lower frequencies, $\omega \ll \omega_0$, one observes 
a power-law growth of $S(\omega)$ 
that corresponds to the persistent time correlations associated
with the ballistic excursions. At even smaller frequencies there is a
crossover to $S(\omega)=$const$=D$, indicating full
decorrelation and pure diffusion.

\begin{figure}
\begin{center}
\includegraphics[width = 8cm]{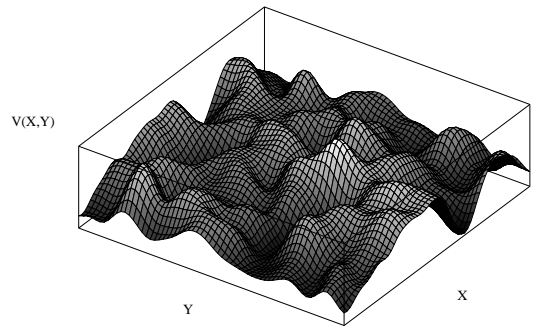}
\end{center}
\caption{
Random potential of the same average potential height as in periodic case. 
}
\label{fig7}
\end{figure}

Disorder in surfaces occurs due to the presence of vacancies and
other defects.  We have generated a random potential surface described by
a Gaussian distribution with a correlation function 
$\langle V({\bf x}) V ({\bf x'})\rangle = 
(\varepsilon/2\pi \lambda'^2)\exp(-|{\bf x}-{\bf x'}|^2/2\lambda'^2)$
(see~\cite{ojalvo,theirvarious,ourvarious,romero2} for details).
We set the intensity $\varepsilon=100$ and the characteristic
length $\lambda'=4$ in our simulations.  A typical surface 
generated with this algorithm whose average height $V_0$ equals
that of the periodic potential is shown in Fig.~\ref{fig7}.

\begin{figure}
\begin{center}
\includegraphics[width = 8cm]{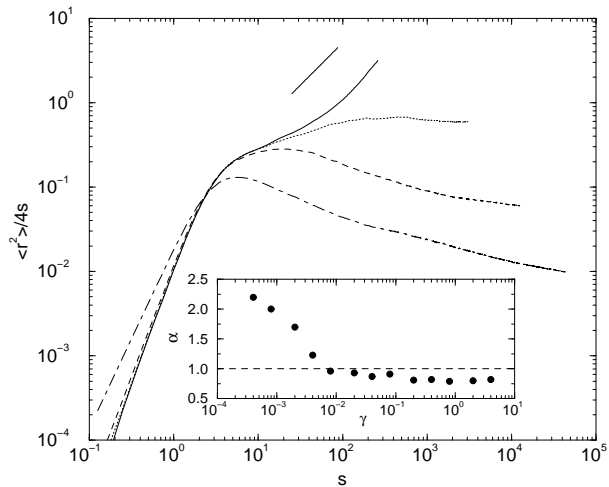}
\end{center}
\caption{
$\langle r^2\rangle/4s$ for a particle in the random potential
for $\gamma=0.0004$ (solid), $0.008$ (dotted),
$0.04$ (dashed) and $0.8$ (dot-dashed).
The straight-line segment has unit slope as a guide to the eye.
Inset: Exponents $\alpha$ versus friction coefficient $\gamma$. 
}
\label{fig9}
\end{figure}

The exponent $\alpha$ in
$\langle r^2(s)\rangle \sim s^\alpha$ at intermediate times
shows the entire range of behaviors \emph{from subdiffusive to
superdiffusive} with
changing friction. In Fig.~\ref{fig9} we show $\langle r^2(s)\rangle/4s$,
averaged over $5000$ particles, as a function of time, for several 
values of $\gamma$.
In the overdamped regime we clearly observe \emph{subdiffusive} behavior 
($\alpha<1$)~\cite{romero2}, while \emph{superdiffusive}
($\alpha>1$) behavior is seen for very small $\gamma$.
The exponents $\alpha$, calculated over the last decade of the time
variation of the mean square displacement within our finite simulation
times, are plotted in the inset of Fig.~\ref{fig9} as a function of $\gamma$. 
Although the subdiffusive behavior is probably the true 
asymptotic behavior in the overdamped case~\cite{romero2}, further theoretical
and numerical support are needed to assess whether superdiffusion is
the asymptotic behavior in the underdamped case.

We summarize our findings. We have explored the behavior of a particle 
in a two-dimensional potential described by ordinary Langevin dynamics
under conditions of thermal equilibrium. In a
periodic potential, in the underdamped regime, the motion of the
particle includes a ballistic range that can extend over many decades of
time. The pdf of the particle's displacements under these conditions
shows a structure strongly resembling one for L\'evy walks. 
This may explain a number of observations involving superdiffusive
motion of organic molecules~\cite{prl1} and atomic clusters~\cite{prl2}
on surfaces without the
need to invoke extraordinary fluctuations beyond the usual thermal
description.  The long-time behavior is
diffusive in all cases, and we have been able to predict theoretically the
dependence of the diffusion coefficient on friction over essentially
the entire range of values of the friction parameter \emph{with no
adjustable parameters}.  The situation 
in a random potential is even more complex, and exhibits  a wide range of 
subdiffusive to superdiffusive regimes. Further
analysis of
the random potential case, and a more extensive presentation of the periodic
problem, will be detailed elsewhere~\cite{we}. 

This work was supported by the MCyT (Spain) under project BFM2003-07850,
by the Engineering Research Program of
the Office of Basic Energy Sciences at the U. S. Department of Energy
under Grant No. DE-FG03-86ER13606, and by a grant from the University of
California Institute for
M\'exico and the United States (UC MEXUS) and the Consejo Nacional de
Ciencia y Tecnolog\'{\i}a de M\'{e}xico (CoNaCyT).
A.H.R. acknowledges support from {\it Millennium Initiative,}
Conacyt-Mexico, under Grant W-8001.
I.M.S. acknowledges the hospitality of the University of Barcelona 
under the CEPBA grant, as well as partial financial support by 
the Fonds der Chemischen Industrie.

\end{document}